\documentclass[twocolumn,amsmath,reprint]{revtex4-1}

\usepackage{amsthm}
\usepackage{amsmath}
\usepackage{amssymb}
\usepackage{caption}
\usepackage{graphicx}
\usepackage{url}
\usepackage{dsfont}
\usepackage{float}
\usepackage{bm}
\usepackage{ifthen}
\usepackage[usenames,dvipsnames]{color}
\usepackage{mathrsfs}
\usepackage[colorlinks=true,citecolor=blue,urlcolor=black]{hyperref}
\usepackage{float}

\newcommand{\be}{\begin{equation}}
\newcommand{\ee}{\end{equation}}
\newcommand{\com}[1]{{\color{Black} #1}}

\usepackage{sistyle}

\newcommand{\mean}[1]{\langle #1 \rangle}

\newcommand{\sket}[1]{{\ensuremath{\lvert#1\rangle}}}
\newcommand{\lket}[1]{{\ensuremath{\left\lvert#1\right\rangle}}}
\newcommand{\ket}[1]{\if@display\lket{#1}\else\sket{#1}\fi}

\newcommand{\sbra}[1]{{\ensuremath{\langle#1\rvert}}}
\newcommand{\lbra}[1]{{\ensuremath{\left\langle#1\right\rvert}}}
\newcommand{\bra}[1]{\if@display\lbra{#1}\else\sbra{#1}\fi}

\newcommand{\sbraket}[2]{{\ensuremath{\langle#1\rvert#2\rangle}}}
\newcommand{\lbraket}[2]{{\ensuremath{\left\langle#1\!\left\rvert\vphantom{#1}#2\right.\!\right\rangle}}}
\newcommand{\braket}[2]{\if@display\lbraket{#1}{#2}\else\sbraket{#1}{#2}\fi}

\newcommand{\sketbra}[2]{{\ensuremath{\lvert #1\rangle\!\langle #2\rvert}}}
\newcommand{\lketbra}[2]{{\ensuremath{\left\lvert #1\right\rangle\!\!\left\langle #2\right\rvert}}}
\newcommand{\ketbra}[2]{\if@display\lketbra{#1}{#2}\else\sketbra{#1}{#2}\fi}

\begin{document}

\title{Quantum random number generation for 1.25 GHz quantum key distribution systems}

\title{Quantum random number generation for 1.25 GHz quantum key distribution systems.}
\author{ A. Martin}
\email{anthony.martin@unige.ch}
\author{B. Sanguinetti}
\author{C. C. W. Lim}
\author{R. Houlmann}
\author{H. Zbinden}

\affiliation{Group of Applied Physics, University of Geneva, Switzerland}


\begin{abstract}
\com{Security proofs of quantum key distribution (QKD) systems usually assume that the users have access to source of perfect randomness.}
State-of-the-art QKD systems run at frequencies in the GHz range, requiring a sustained GHz rate of generation and acquisition of quantum random numbers. In this paper we demonstrate such a high speed random number generator. The entropy source is based on amplified spontaneous emission from an erbium-doped fibre, which is directly acquired using a standard small form-factor pluggable (SFP) module. The module connects to the Field Programmable Gate Array (FPGA) of a QKD system. A real-time randomness extractor is implemented in the FPGA and achieves a sustained rate of 1.25 Gbps of provably random bits.
\end{abstract}

\maketitle

\section{Introduction}
Proposed in 1984, quantum key distribution (QKD) is a cryptographic technique that allows two spatially separated users, called Alice and Bob, to exchange secret key via a potentially insecure quantum channel. A key advantage of QKD is that it offers provably secure cryptographic keys, while classical key distribution schemes can only offer computational security. Nowadays, owning to the significant progress made in photonics technology, QKD experiments are able to distribute large secret keys at high repetition rates~\cite{Liu2010,Wang2012,Walenta2014,Patel2014}. Among the many developments required for a fast QKD system, is the development of a fast and practical random number generator where the randomness is derived from quantum processes~\cite{Kanter2009,Li2010a,Li2011a,Wahl2011,Symul2011,Zhang2012,Nguimdo2012,Xu2012,Lozach2013,Abellan2014}. However, the integration of these fast quantum random number generators (QRNGs) with fast QKD systems has remained elusive, mainly due to the technical challenges in acquiring high data rates and certifying that the generated numbers are truly random. Moreover, the development of random number generator based on physical source of randomness is useful for the classical cryptography protocols too. All the recent social events have demonstrated that cryptography protocols using a pseudo-random number generators is not at all secure~\cite{Markowsky2014}.

In this paper, we demonstrate how a fast quantum random number generator can be easily created and integrated into a communication system. In our device, random numbers are generated by a source of amplified spontaneous emission (ASE) and acquired by a field-programmable gate array (FPGA) using a standard small form-factor pluggable (SFP) module. This allows us to generate true random numbers at a rate of \SI{1.25}{Gbit/s}, as required by state-of-the-art QKD systems. In the following, we first describe the physical origin of the randomness that our random number generation is based on. Then, we sketch the experimental setup and detail how each element affects the randomness of our data-stream. Finally, we present our results, the amount of generated entropy and how it can be efficiently extracted.

\section{Setup of random number generator}

\begin{figure}
\begin{center}
\includegraphics[width=\columnwidth]{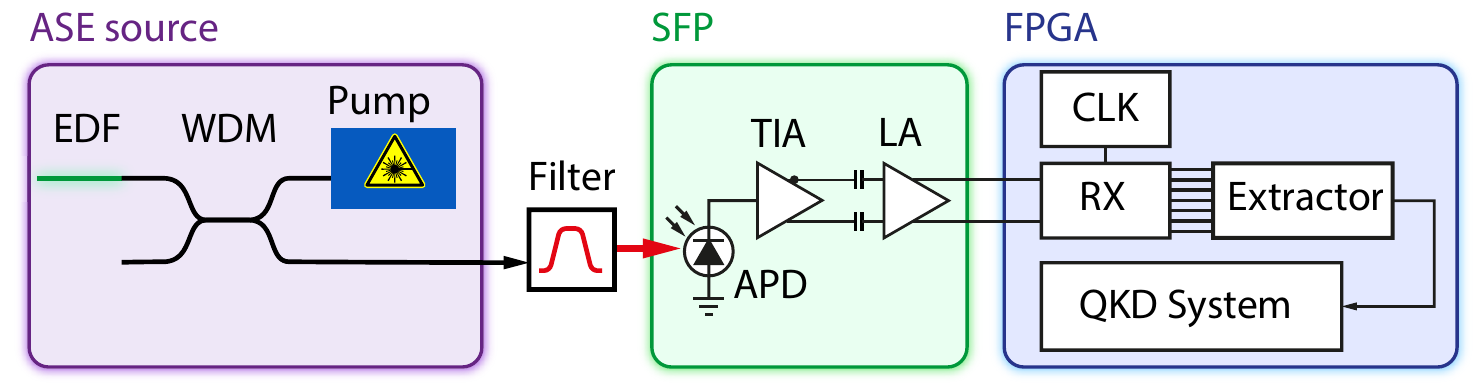}
\caption{Experimental setup, composed of an amplified spontaneous emission (ASE) source, a filter, an SFP module and an FPGA. The ASE source consists of a back-pumped erbium doped fibre (EDF) emitting at \SI{1530}{nm}. The emitted light is filtered with a bandpass filter (\SI{50}{GHz} DWDM) and sent to the SFP module. This module consists of an avalanche photodiode (APD), which is further amplified by a transimpedance amplifier (TIA). The signal is then ``digitised'' by a limiting amplifier (LA) and sent on as a differential pair to the FPGA.  The FPGA's receiver (RX) deserializes the bitstream, using a stable sample clock (CLK) and passes it to the randomness extractor. Finally, the extracted numbers can be used by the quantum key distribution system.}
\label{fig_ASE_RNG_schematic}
\end{center}
\end{figure}

The experimental setup is simple and compact, in that it is constructed with standard telecommunications components, as shown in \figurename{~\ref{fig_ASE_RNG_schematic}}. The randomness source is based on a filtered ASE source followed by a small form-factor pluggable (SFP) module plug to a field-programmable gate array (FPGA). Below we describe these elements, analyzing the sources of randomness and of error.

\subsection{Source ASE}

The generation of quantum random numbers using ASE has been described in detail in~\cite{Williams2010}. Here we briefly review this method, for an ASE source based on a pumped Erbium doped fibre.

\begin{figure}[htbp]
\begin{center}
\includegraphics[width=\columnwidth]{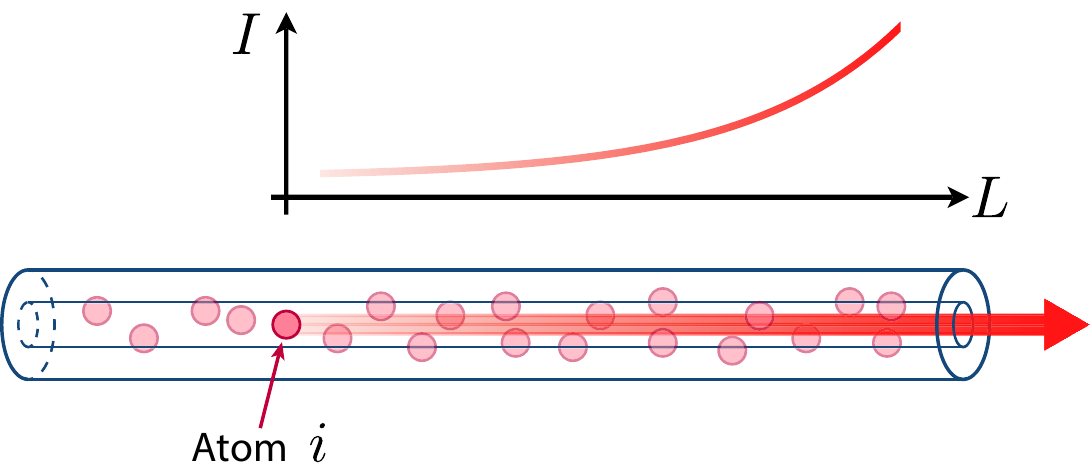}
\caption{In an erbium-doped fibre, atoms have an equal probability of emitting a photon at any point along the fibre. This photon will be amplified exponentially with the distance travelled before exiting the doped fibre.}
\label{fig_EDF}
\end{center}
\end{figure}

Erbium doped fibers are employed to realize telecommunication wavelength amplifiers. For that, a pump laser at 980\,nm excites the erbium atoms, which are homogeneously distributed over the fiber. This constitutes a gain medium, such as used in erbium-doped fibre amplifiers (EDFAs).
Each excited atoms has an equal probability of decaying. As shown in Fig.~\ref{fig_EDF}, depending on the position of this atom along the fibre, the photon will travel a length $L$ in the gain medium before it reaches the end of the doped fibre. During this journey, the photon will stimulate the emission of a number of photons proportional to the exponential of $L$. This results in a very large distribution of output powers, described by Bose-Einstein statistics:
\begin{equation}
P_\text{BE}(\bar{n},n) := \frac{\bar{n}^n}{(1+\bar{n})^{1+n}},
\label{eq:thermal_distribution}
\end{equation}
where $P_\text{BE}(\bar{n},n)$ is the probability of finding $n$ photons in one mode which is populated by $\bar{n}$ photons on average~\cite{Wong1998}.
%

\begin{figure}[htbp]
\begin{center}
\includegraphics[width=\columnwidth]{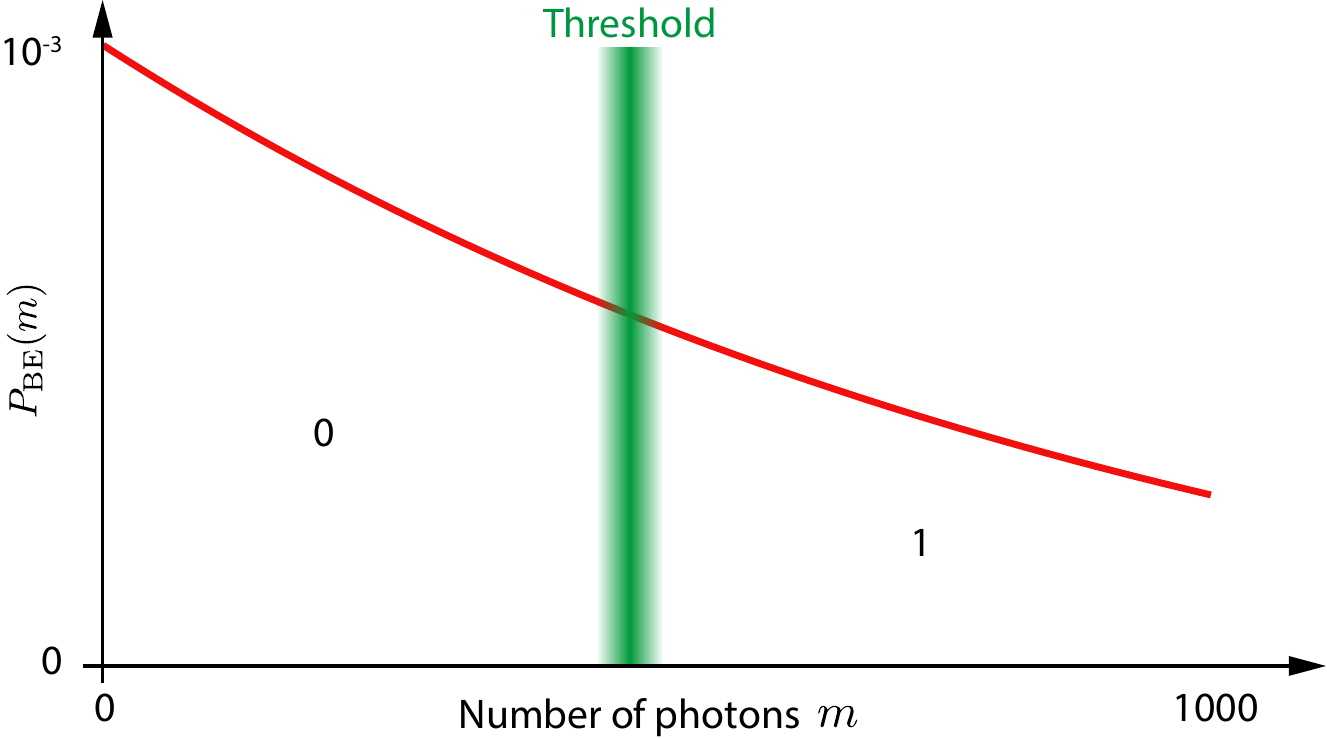}
\caption{Bose-Einstein photon number distribution for a single-mode field. This distribution is large, so that when discriminated by a threshold, the generated bit,  $0$ or $1$, will depend on this distribution rather than on electrical noise.}
\label{fig_ASE_thermal_state}
\end{center}
\end{figure}

\begin{figure}[htbp]
\begin{center}
\includegraphics[width=\columnwidth]{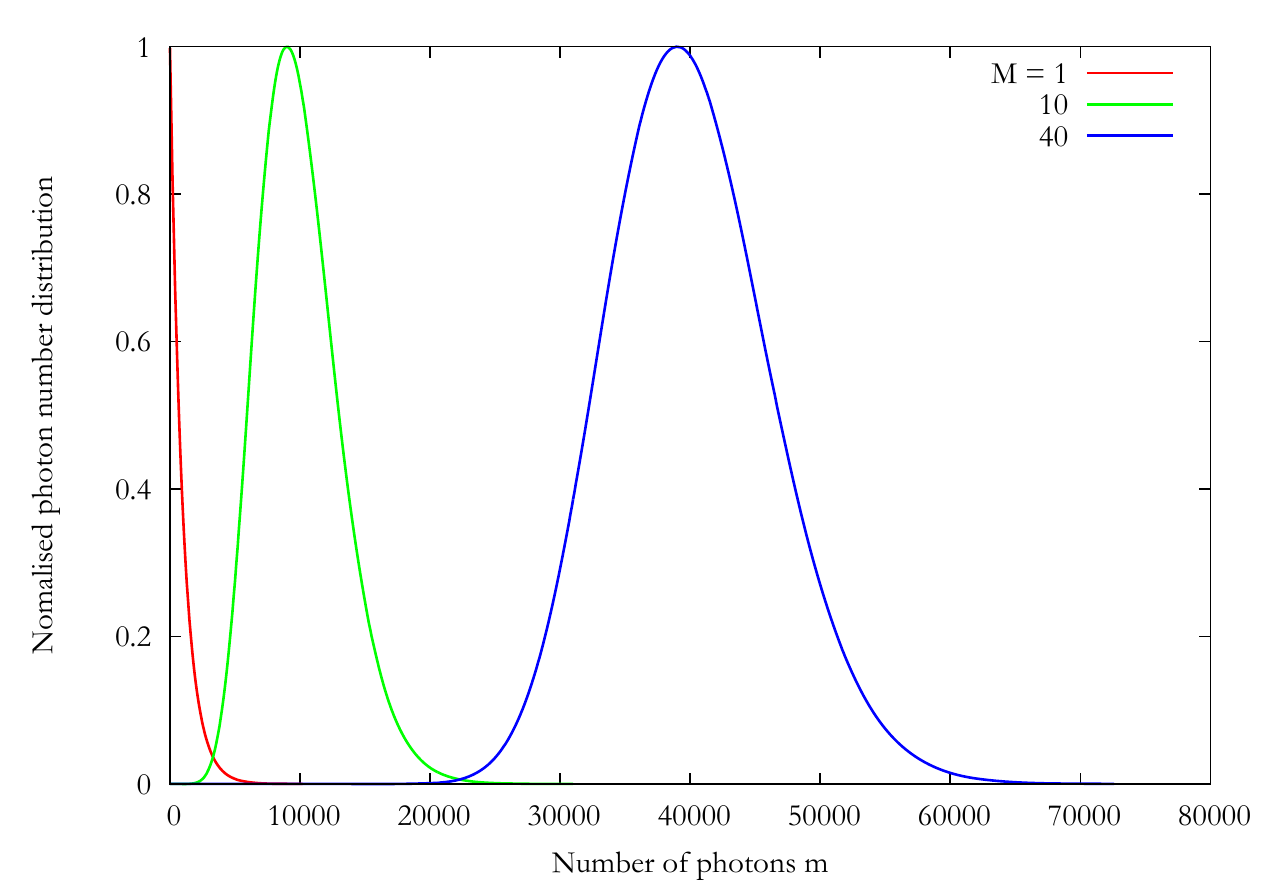}
\caption{Normalized photon number distribution of a thermal state with a mean photon number of 1000, and a number of modes $M=1,10,40$ respectively. As the number of modes grows, so does the standard deviation of the distribution. A larger number of modes also results in a more symmetric distribution.}
\label{fig_ASE_multimode_thermal}
\end{center}
\end{figure}

\figurename{~\ref{fig_ASE_thermal_state}} shows this distribution for a mean number $\bar{n}=\SI{1000 }{photons/mode}$. This distribution is broad, so that a random number can be generated by looking whether the instantaneous intensity is above or below a specific threshold. To certify that the collected entropy is mostly of quantum origin, it is important that the photocurrent distribution is much broader than the distribution due to electrical (classical) noise.
%
 A practical source of ASE will output approximately $\bar{n}=\SI{1000}{photons/mode}$ equivalent to ~\SI{600}{nW} in a \SI{2.5}{GHz} bandwidth, or \SI{-32}{dBm}, which is close to the noise  level of receivers using an avalanche photodiode (APD). It is possible to send a higher number of modes onto the APD, \textit{i.e.} choose an optical signal with a spectral bandwidth larger than the detector bandwidth by a factor $m$. The average number of photons will then grow by a factor $m$, and the standard deviation, which is the value of interest to us, will grow by a factor $\sqrt{m}$. In this case, the intensity distribution is described by~\cite{Wong1998}:
\begin{equation}\label{Eq_stat_mode}
P_\text{BE}(\bar{n},n,m) = \frac{\Gamma (n+m)}{\Gamma(n+1)\,\Gamma(m)}\left(1+\frac{1}{\bar{n}}\right)^{-n} (1+\bar{n})^{-m}
\end{equation}
This distribution is plotted in \figurename{~\ref{fig_ASE_multimode_thermal}} for a number of modes $m=1,10\text{ and } 40$. As shown in this graph, using a higher number of modes makes the distribution more symmetric, which reduces bias, and increases it's standard deviation, which makes detection easier.

For our source, we employed 1\,m of erbium doped fibre (Thorlabs: ER30-4/125). The 980\,nm pump laser output of \SI{600}{mW} is sufficiently strong to saturate the medium, which allows to achieve a spectral radiance of the order of 1000 photons/mode ,equivalent to that of a 30dB amplifier, and equivalent to $\sim\SI{10}{\micro W}$ in a \SI{50}{GHz} bandwidth. The spectral radiance of the source was calibrated using a fibre-based radiometer~\cite{Lunghi2014}, and is perfectly stable on a long timescale (milliseconds to hours)~\cite{Monteiro2013}.

\subsection{Detection}

Detection is carried out by the APD receiver of an SFP (FTRJ1621P1BCL) module. As presented in \figurename{~\ref{fig_ASE_RNG_schematic}}, the InGaAs APD cathode is directly connect to a trans-impedance amplifier (TIA). At the output of the TIA, the signal is AC coupled to a limiting amplifier (LA) which generate the data coded  as a voltage, + 400 mV and - 400 mV corresponding to a bit 1, and 0, respectively. More precisely, for a signal greater (smaller) than the average the LA delivers a voltage of 400\,mV (-400\,mV). This signal could be directly plug to an FPGA to perform the post-processing.

The module works at a bandwidth of 2.5 GHz, which fixes at 40 the number of modes received by the detectors. With equation (\ref{Eq_stat_mode}), the standard deviation of the number of photons arriving on the sensor is $\sigma_p = 6246$, with a mean value of 40000 (see \figurename{~\ref{fig_ASE_multimode_thermal}}). The equivalent photon noise introduced by the detector and the electronics is $\sigma_n = 610$, which is one order of magnitude smaller than the standard deviation of the light.

\section{Characterization and results}

\begin{figure}
\begin{center}
\includegraphics[width=\columnwidth]{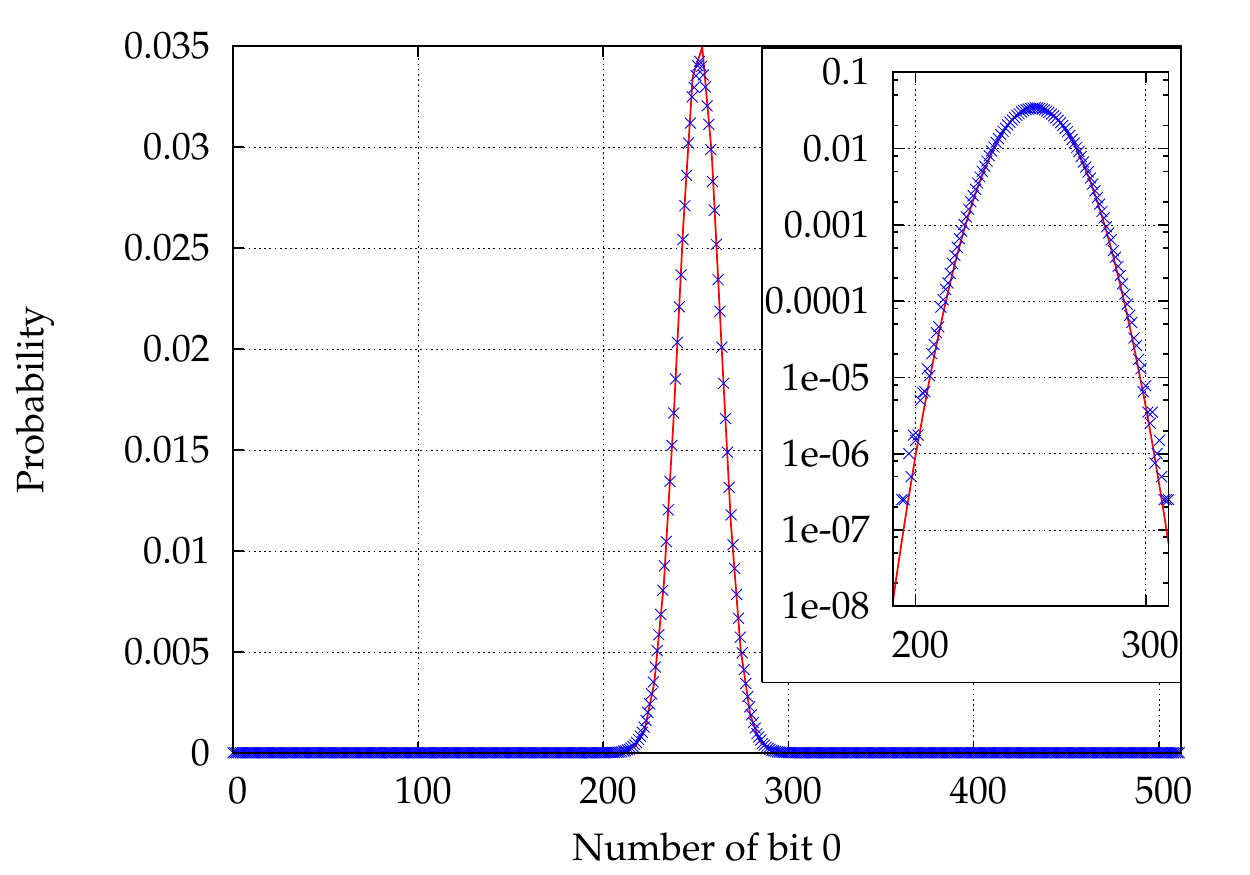}
\caption{\label{fig_P0}Histogram representing the probability to obtain zeros in a block of 512 element for $4\times 10^6$ blocks. The red curve represent the statistical spread due to the small sample size predict by the equation (\ref{eq_binomial}). The insert represents the same curves in log scale.}
\end{center}
\end{figure}

First, we estimate the probability of obtaining a zero or a one. In our configuration, the SFP automatically sets the threshold for the discrimination with an AC coupled limiting amplifier. To ensure that the discriminator does not move during the time and consequently change the statistic of our RNG, we acquired $4\times10^6$ blocks of 512 successive bits over a period of two days. For each block, the number of bits 0 are counted, to construct the probability histogram reported in \figurename{~\ref{fig_P0}} (blue dots). More precisely, we expect the probability to obtain k zeros in a string of $n$ bits to be given by the Binomial distribution :
\begin{equation}\label{eq_binomial}
P(k) = \frac{n!}{k!(n-k)!} {P_0}^k (1-P_0)^{n-k},
\end{equation}
where $P_0$ the probability to obtain a bit 0. This function is represented by a red line in the \figurename{~\ref{fig_P0}}, with $P_0 = 0.4920$ given by the mean value of the measured data. The standard deviation of the function is given by $\sigma_T = \sqrt{nP_0(1-P_0)} = 11.31$, which is close to the measured value $\sigma_M = 11.68(1)$. \com{Roughly speaking, this shows that the underlying distribution is well approximated by a Binomial distribution. In conclusion, the probability to obtain 0 or 1 is not equiprobable---this is due to the slight misalignment on the threshold position. However, note that the threshold position is stable over the acquisition time. Indeed, if the threshold position is not stable, or if the power of the source changes during the acquisition time, then this may lead to a broader distribution.}

\begin{figure}
\begin{center}
\includegraphics[width=\columnwidth]{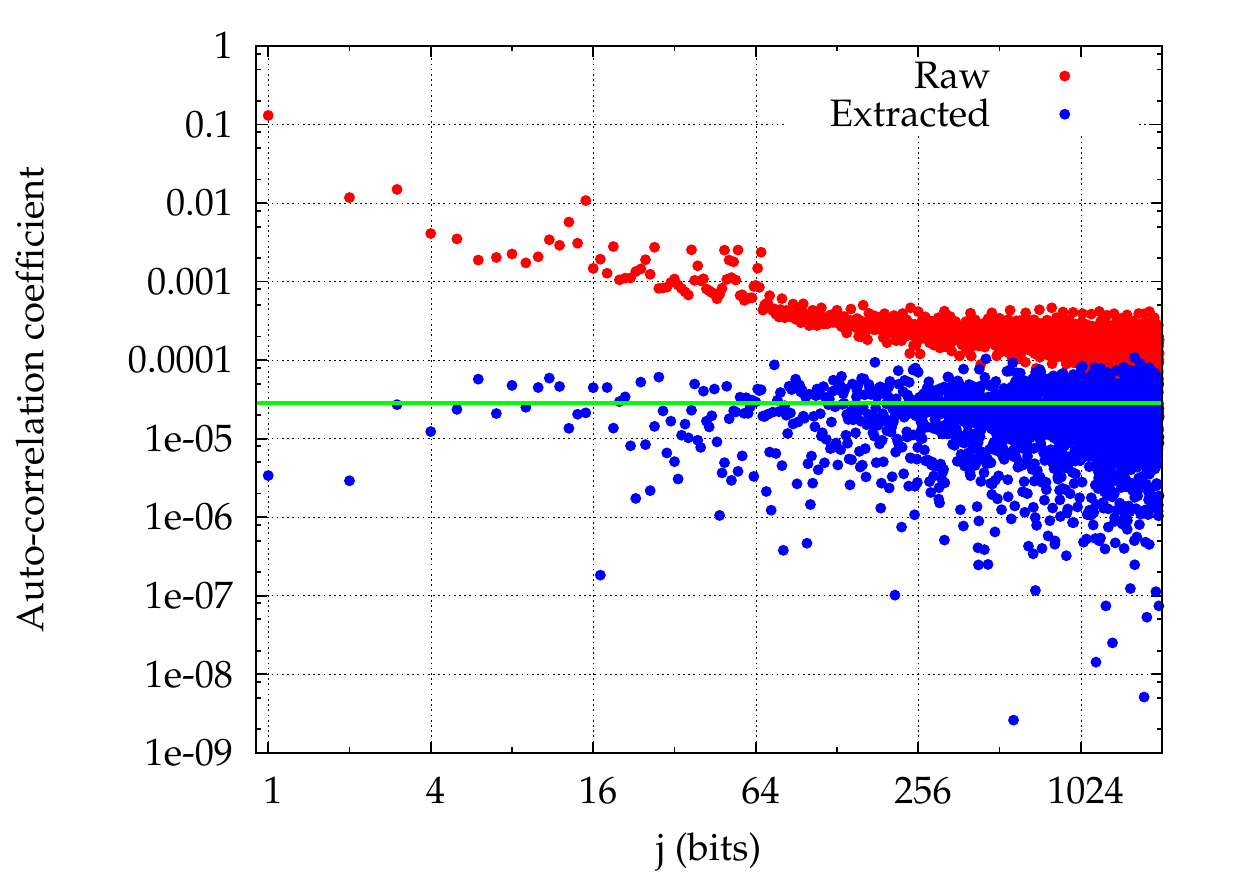}
\caption{\label{fig_xcorr} Auto-correlation coefficient computed for a raw and extracted sequences of 1.25\,Gbits. The green line represent the standard deviation of the correlation due to the finale size of the sample.}
\end{center}
\end{figure}

Next, we compute the autocorrelation function, which measures the serial correlation within a data sample. 
\begin{equation}
R(j) =\frac{ \mean{X_0X_j}- \mean{X_0}\mean{X_j}}{\sqrt{\left(\mean{X_0^2}-\mean{X_0}^2\right)\left(\mean{X_j^2}-\mean{X_j}^2\right)}},
\end{equation}
where $\mean{\cdot}$ denotes the proportion of ones, and $X_0$ and $X_j$ represent a bit string X[0,n] and X[j,n+j], respectively. The red curve in \figurename{~\ref{fig_xcorr}} represent this function obtain for a string of $1.25\times10^9$ successive bits. The correlations are not negligible, therefore we cannot model the data with an iid (independent and identically distributed) process. \com{To get a bound on the min-entropy of the source, we consider a Santha-Vazirani type of weak source of randomness~\cite{Santha1986,Plesch2013}, which generates a sequence of weak random bits where the probability distribution of each bit, conditioned on the previously generated bits, is bounded by some fixed real parameter $0<\delta\leq 1/2$. In particular, let the $X_1, X_2, \ldots, X_n$ be a sequence of $n$ random bits generated by the Santha-Vazirani source, then for each $0\leq i < n$
\begin{equation}\label{eq_SVsource}
\delta \leq P(X_{i+1}=x_{i+1} | X_i=x_i,\dots,X_1=x_1) \leq 1-\delta.
\end{equation}
For this type of source, the min-entropy is bounded by $H_{\rm min} (X_1\ldots X_n)\geq -{\rm log_2}((1-\delta)^n)$. Here, it is important to note that the min-entropy measures the number of extractable random bits from the sequence of $n$ weakly random bits.}

One approach to estimate the value of $\delta$ consists in using the correlation coefficient. The stronger correlation is obtained for j = 1 with R(1) = 0.13078 and all the other is at least one order of magnitude smaller. With the correlation coefficient, we calculate the mutual information given by $I(X_1,X_2) = \frac{1}{2} \ln \left( 1-R(1)^2\right)$, and the conditional entropy $H(X_1|X_0) = H(X_0) - I(X_0,X_1) = 0.9912$ for $P_0$ estimate in the previous section. To estimate the min-entropy, we compute the greatest conditional probability satisfying the conditional entropy defined in previous equation, which corresponds to $max(P(x_1|x_0)) = 0.5775$ and $\delta = 0.4225$. This approach allows to correctly estimate the bound of the min-entropy for n=2, but for $n\geq2$ we assume that only two successive bits are correlated.

\begin{figure}
\begin{tabular}{l}
a) \\
\includegraphics[width=\columnwidth]{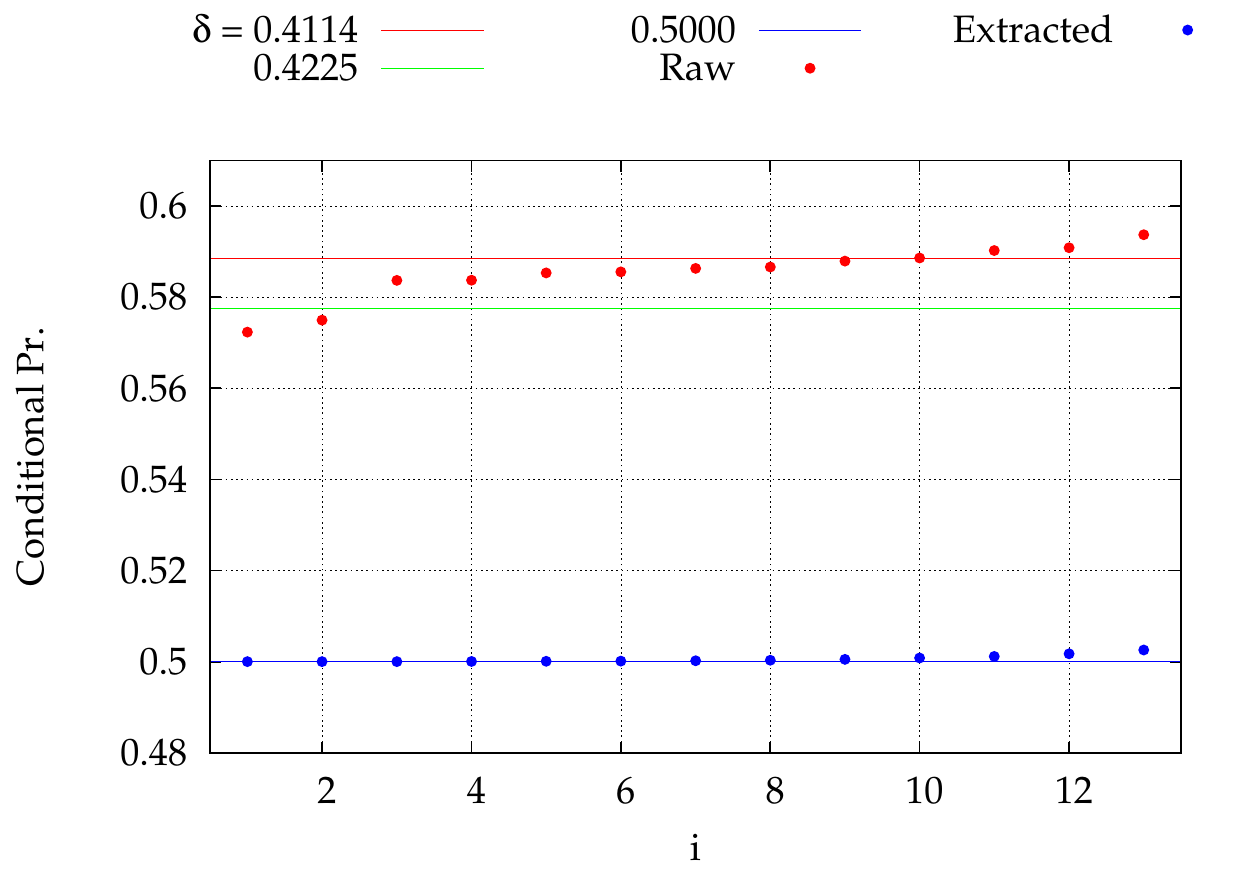} \\
b) \\
\includegraphics[width=\columnwidth]{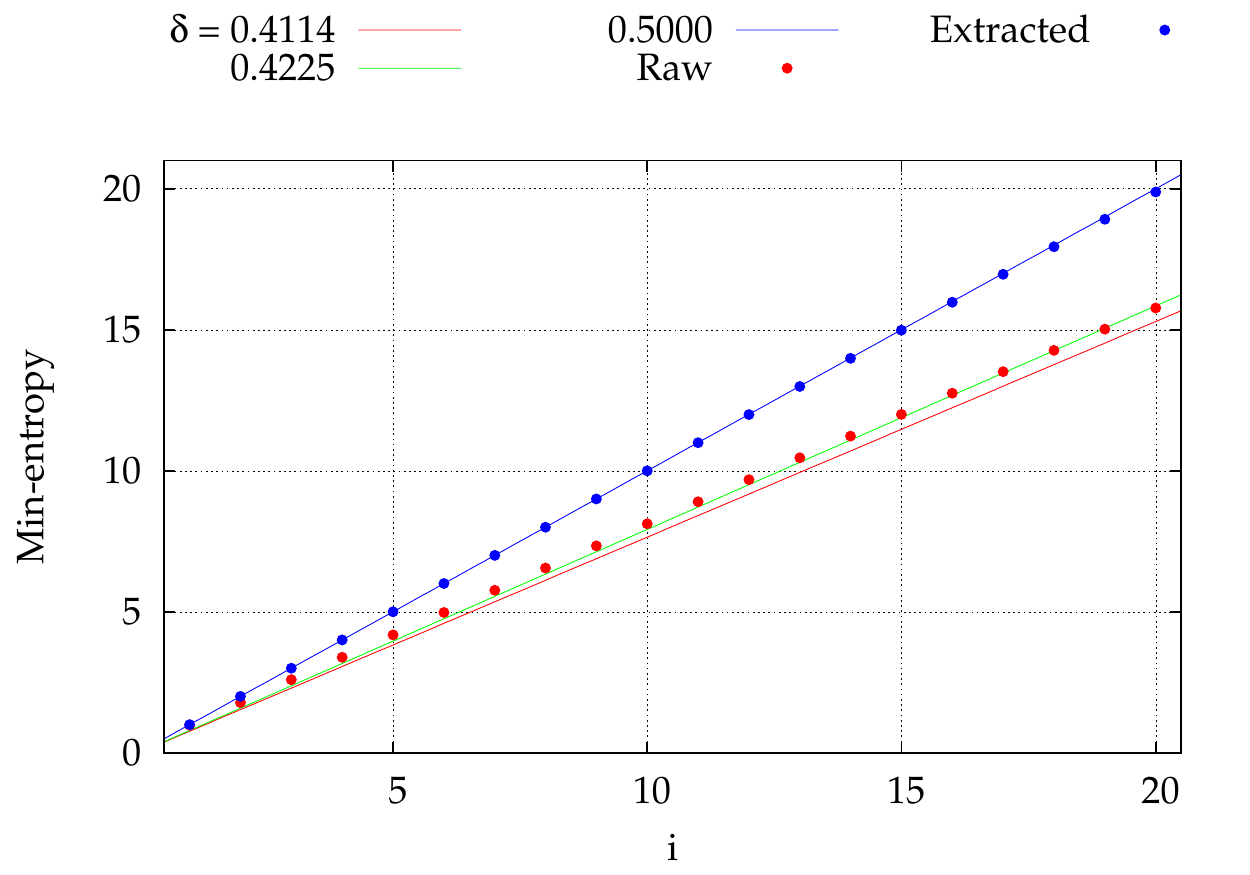}
\end{tabular}
\caption{\label{fig_delta} a) Maximal conditional probability measures from $5\times 10^9$ bits as the function of the number of known bits. b) Min-entropy as a function of the string length i. }
\end{figure}

To improve the estimation of $\delta$, we can calculate the conditional probability when we know more bits, which is given by
\begin{align*}
P(X_i&=x_i | X_1=x_1 ... X_{i-1}=x_{i-1}) \\
 &= \frac{P(X_1=x_1 ... X_{i-1}=x_{i-1},X_i=x_i )}{P(X_1=x_1 ... X_{i-1}=x_{i-1})}.
\end{align*}
The maximum conditional probability as function of i obtained by this method is shown in \figurename{~\ref{fig_delta}~a)}. These values are calculated from a string of $5\times10^9$ bits. We can remark that the estimation of $\delta$ with the previous method is not accurate for $i\geq2$. Furthermore, with this approach, for $i$ large, the error introduced by the finite size of the sample used to estimate the conditional probability are no longer negligible. More precisely, for i larger than 10 the results are not conclusive. For i=10, we obtain $\delta = 0.4114$.

To verify if the estimation the estimation of delta is correct, we compute the min-entropy of 5~Gbit for different string size $n$, i.e. $H_{\rm min} (X_1\ldots X_n)= -{\rm log_2}(\max(P(X_n ... X_0))$, as show the \figurename{~\ref{fig_delta}~b)}. Moreover, the lines represent the min-entropy value for two different $\delta$. For $\delta = 0.4225 $, $\mathcal{K}$ is greater than the computed $H_{\rm min}$ the large n, which is not the case for $\delta = 0.4114$. As shown \figurename{\ref{fig_xcorr}}, the correlation coefficient after 20 bits are the order of $10^{-3}$, we could assume the $\delta = 0.4114$ verify the equation (\ref{eq_SVsource}) $\forall n$.

To improve the min-entropy per bit, an extractor is implement in the FPGA based bit-matrix-vector multiplication with a random matrix\cite{troyer2012randomness,Frauchiger2013}. The extractor uses raw sequence of 512 bits to generate 256 extracted bits, by multiplying the raw sequence with the random matrix of size $512\times256$ generated by a QRNG Quantis. In principle, with the min-entropy of the physical random generator, we could generated at least $391$ bits per cycle. We choose to generate only 256 bits per cycle to obtain a frequency of 1.25\,GHz. Moreover, a greater compression factor ensure the generation of a near perfect random bit string.

\begin{figure}
\begin{center}
\includegraphics[width=\columnwidth]{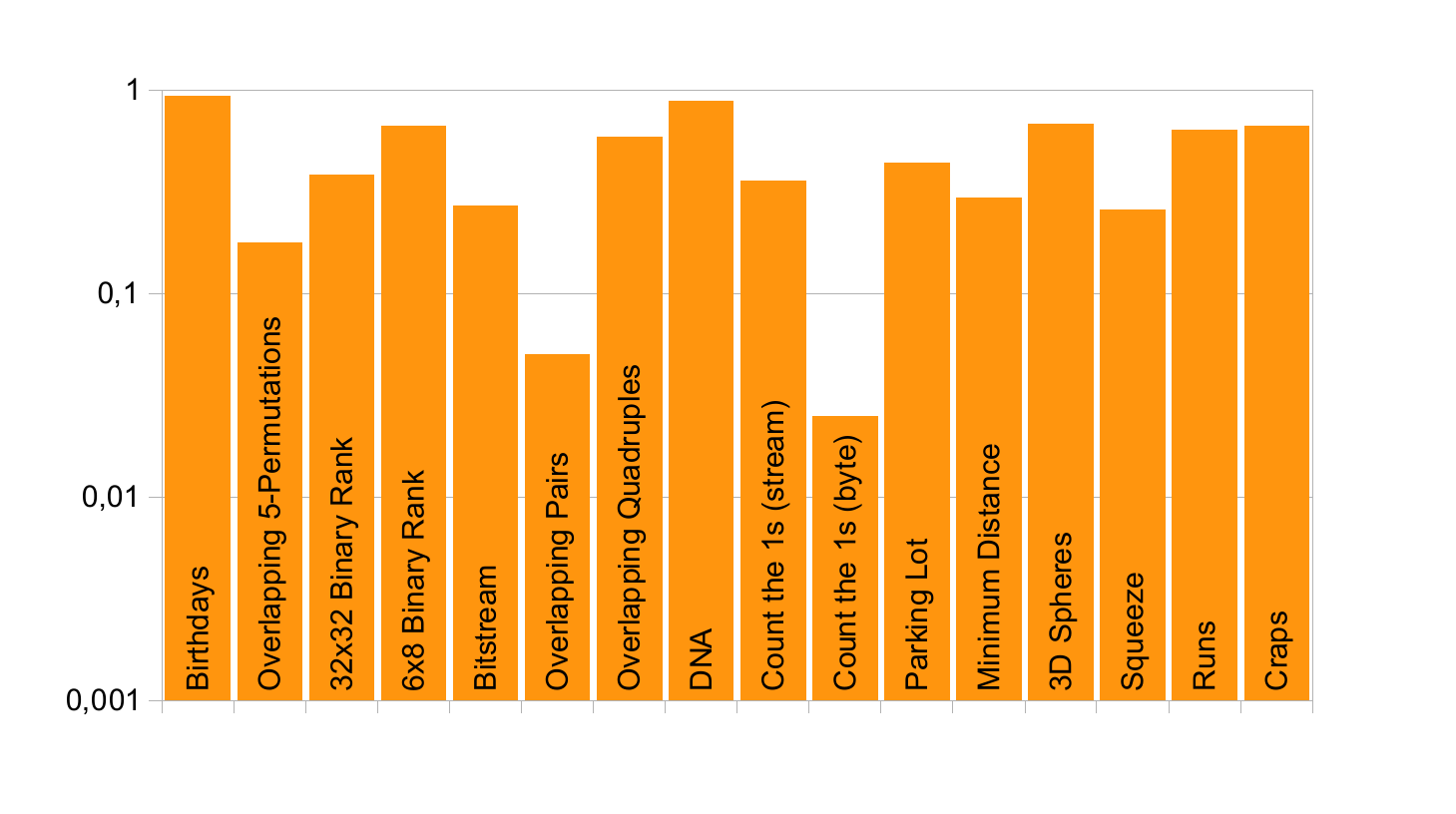}
\caption{\label{fig_dieharder}Dieharder test of the data at the output of the extractor. For each test the represented p-value is the result of a Kolmogorov-Smirnov test of 100 p-values. To pass the test the p-value needs to satisfy $0.01 \leq p \leq 0.99$~\cite{Alani2010}.}
\end{center}
\end{figure}

To characterize the randomness of the extracted bit string, we first calculated the probability to obtain a bit 0 after 1 s of measurement, $P^e_0=0.500001 (15)$. Then, we computed as for the raw data the autocorrelation coefficient, the conditional probability and the min-entropy (see \figurename{~\ref{fig_xcorr} and \ref{fig_delta}}). As shows all these measurements confirm the good randomness of the bits after the extractor. Moreover, the \figurename{~\ref{fig_dieharder}} present some test results obtained with the extracted data. Note that, all the Diehard tests are successfully passed.

\section{Conclusion}

In this paper we present a random number generator based on the statistics of a ASE source implemented in an erbium doped fibre. In addition to the completely fibered source of light, our generator takes advantage of a SFP module and an FPGA which makes compact, intrinsically stable, and standalone. Indeed, the implementation of all the post-processing inside the FPGA allows us to generate a random data flow of 1.25 Gb/s  in real time. Our proposal has the benefit that it can be easily upgraded to a higher throughput, e.g., \SI{10}{Gbit/s} simply by using a faster ``SFP+'' module.

This generator is an excellent candidate to generate the random numbers necessary during the implementation of commercial quantum key distribution based on protocols such as COW or BB84.

\section*{Acknowledgement}
This work was supported by the Swiss NCCR QSIT.

\bibliography{Paper_Submission}

\end{document}